\documentclass[11pt,a4paper]{article}

\newif\ifplainstyle
\plainstylefalse

\newif\ifjhepstyle
\jhepstyletrue

\newif\ifprstyle
\prstylefalse

\ifjhepstyle
	\usepackage{jheppub}
	\usepackage{amsfonts}
	\usepackage{verbatim}
	\usepackage{float}
	\usepackage{array}
	\newcolumntype{C}[1]{>{\centering\arraybackslash$}p{#1}<{$}}
	\makeatletter
	\def\@fpheader{\phantom{Prepared for submission to JHEP}}
	\makeatother
\else	
 	\ifprstyle
		\usepackage{verbatim}
		\usepackage{amsmath,amsfonts,amssymb}
		\usepackage[colorlinks=true
                	,urlcolor=blue
                	,anchorcolor=blue
                	,citecolor=blue
                	,filecolor=blue
                	,linkcolor=blue
                	,menucolor=blue
                	]{hyperref}
	\else
            	\usepackage{verbatim}
            	\usepackage{cite}
            	\usepackage{setspace}
            	\usepackage[top=2.5cm, bottom=2.75cm, left=2.5cm, right=2.5cm]{geometry}
            	\usepackage{amsmath,amsfonts,amssymb}
            	\usepackage[colorlinks=true
            	,urlcolor=blue
            	,anchorcolor=blue
            	,citecolor=blue
            	,filecolor=blue
            	,linkcolor=blue
            	,menucolor=blue
            	,linktoc=page
            	]{hyperref}
            	\usepackage{float}
            	\restylefloat{table}
            	
            	\numberwithin{equation}{section}
	\fi
\fi

\allowdisplaybreaks

\newcommand{\ThisIsTheTitle}{The covariant action of higher spin black holes in three dimensions}
\newcommand{\ThisIsAuthorOne}{Luis Apolo}
\newcommand{\ThisIsEmailOne}{luis.apolo@fysik.su.se}
\newcommand{\ThisIsTheAffiliation}{Department of Physics \& The Oskar Klein Centre, \\
Stockholm University, AlbaNova University Centre, SE-106 91 Stockholm, Sweden}
\newcommand{\TheseAreTheKeywords}{}

\newcommand{\ThisIsTheAbstract}{We propose a set of boundary terms for higher spin theories in AdS$_3$ that lead to a well-defined variational principle compatible with Dirichlet boundary conditions for the metric and higher spin fields. These boundary terms are valid for higher spin theories in the Fefferman-Graham gauge and they allow us to compute the canonical free energy of higher spin black holes directly from the Euclidean, covariant, on-shell action. Using these results we reproduce the thermodynamics of the higher spin black hole of Ammon, Gutperle, Kraus, and Perlmutter and comment on the corresponding theory of induced $\cal{W}$-gravity at the boundary.}

\ifjhepstyle
\title{\ThisIsTheTitle}

\author{\ThisIsAuthorOne}

\affiliation{\ThisIsTheAffiliation}

\emailAdd{\ThisIsEmailOne}

\abstract{\ThisIsTheAbstract} 

\fi

\begin{document}

\ifjhepstyle
\maketitle
\flushbottom
\fi

\long\def\symfootnote[#1]#2{\begingroup%
\def\thefootnote{\fnsymbol{footnote}}\footnote[#1]{#2}\endgroup} 

\def\({\left (}
\def\){\right )}
\def\lb{\left [}
\def\rb{\right ]}
\def\lB{\left \{}
\def\rB{\right \}}

\def\Int#1#2{\int \textrm{d}^{#1} x \sqrt{|#2|}}
\def\Bra#1{\left\langle#1\right|} 
\def\Ket#1{\left|#1\right\rangle}
\def\BraKet#1#2{\left\langle#1|#2\right\rangle} 
\def\Vev#1{\left\langle#1\right\rangle}
\def\Vevm#1{\left\langle \Phi |#1| \Phi \right\rangle}\def\bbox{\bar{\Box}}
\def\til#1{\tilde{#1}}
\def\wtil#1{\widetilde{#1}}
\def\ph#1{\phantom{#1}}

\def\ra{\rightarrow}
\def\la{\leftarrow}
\def\lra{\leftrightarrow}
\def\p{\partial}
\def\diff{\mathrm{d}}

\def\sinh{\mathrm{sinh}}
\def\cosh{\mathrm{cosh}}
\def\tanh{\mathrm{tanh}}
\def\coth{\mathrm{coth}}
\def\sech{\mathrm{sech}}
\def\csch{\mathrm{csch}}

\def\a{\alpha}
\def\b{\beta}
\def\g{\gamma}
\def\d{\delta}
\def\e{\epsilon}
\def\ve{\varepsilon}
\def\k{\kappa}
\def\l{\lambda}
\def\n{\nabla}
\def\om{\omega}
\def\s{\sigma}
\def\t{\theta}
\def\z{\zeta}
\def\vp{\varphi}

\def\ss{\Sigma}
\def\dd{\Delta}
\def\gg{\Gamma}
\def\ll{\Lambda}
\def\tt{\Theta}

\def\A{{\cal A}}
\def\B{{\cal B}}
\def\cE{{\cal E}}
\def\D{{\cal D}}
\def\F{{\cal F}}
\def\H{{\cal H}}
\def\I{{\cal I}}
\def\J{{\cal J}}
\def\K{{\cal K}}
\def\L{{\cal L}}
\def\O{{\cal O}}
\def\P{{\cal P}}
\def\cS{{\cal S}}
\def\W{{\cal W}}
\def\X{{\cal X}}
\def\Z{{\cal Z}}

\def\we{\wedge}

\def\tilw{\tilde{w}}
\def\tile{\tilde{e}}

\def\zz{\bar z}
\def\xx{\bar x}
\def\xp{x^{+}}
\def\xm{x^{-}}

\def\VirU1{\mathrm{Vir}\otimes\hat{\mathrm{U}}(1)}
\def\VirSL2R{\mathrm{Vir}\otimes\widehat{\mathrm{SL}}(2,\mathbb{R})}
\def\U1{\hat{\mathrm{U}}(1)}
\def\SL2R{\widehat{\mathrm{SL}}(2,\mathbb{R})}
\def\sl2r{\mathrm{SL}(2,\mathbb{R})}
\def\by{\mathrm{BY}}

\def\RR{\mathbb{R}}

\def\tr{\mathrm{tr}}

\def\sint{\int_{\ss}}
\def\dsint{\int_{\p\ss}}
\def\hint{\int_{H}}

\newcommand{\eq}[1]{\begin{align}#1\end{align}}
\newcommand{\eqst}[1]{\begin{align*}#1\end{align*}}
\newcommand{\eqsp}[1]{\begin{equation}\begin{split}#1\end{split}\end{equation}}

\newcommand{\absq}[1]{{\textstyle\sqrt{|#1|}}}
\newcommand{\absqe}[1]{{\sqrt{#1}}}


\ifprstyle
\title{\ThisIsTheTitle}

\author{\ThisIsAuthorOne}
\email{\ThisIsEmailOne}


\affiliation{\ThisIsTheAffiliation}


\begin{abstract}
\ThisIsTheAbstract
\end{abstract}


\maketitle

\fi

\ifplainstyle
\begin{titlepage}
\begin{center}

\ph{.}

\vskip 4 cm

{\Large \bf \ThisIsTheTitle}

\vskip 1 cm

\renewcommand*{\thefootnote}{\fnsymbol{footnote}}

{{\ThisIsAuthorOne}\footnote{\ThisIsEmailOne}}

\renewcommand*{\thefootnote}{\arabic{footnote}}

\setcounter{footnote}{0}

\vskip .75 cm

{\em \ThisIsTheAffiliation}

\end{center}

\vskip 1.25 cm

\begin{abstract}
\noindent \ThisIsTheAbstract
\end{abstract}

\end{titlepage}

\newpage

\fi

\ifplainstyle
\tableofcontents
\noindent\hrulefill
\bigskip
\fi



\section{Introduction}
\label{se:intro}

Higher spin theories in AdS$_3$ share many of the features that make three-dimensional gravity with a negative cosmological constant an interesting toy model~\cite{Deser:1983tn,Deser:1983nh}. Despite lacking local dynamics, the theory possesses boundary degrees of freedom. These are linearized pertubations associated with the non-trivial asymptotic symmetries of the theory~\cite{Brown:1986nw,Henneaux:2010xg,Campoleoni:2010zq,Campoleoni:2011hg,Campoleoni:2014tfa}. The absence of local degrees of freedom implies that all solutions are locally gauge equivalent. Yet the theory admits globally non-trivial black hole solutions~\cite{Gutperle:2011kf,Castro:2011fm,Ammon:2011nk}, generalizations of the BTZ black holes of three-dimensional gravity~\cite{Banados:1992wn, Banados:1992gq}.

The thermodynamics of black holes in anti-de Sitter space are most elegantly studied in the framework of covariant Euclidean quantum gravity~\cite{Gibbons:1976ue,Hawking:1982dh}. That is, via the Euclidean partition function, whose saddle-point approximation yields the black hole free energy $\F$,
  \eq{
  Z[h] = \int \D g \, e^{- I_E[g,h] } \sim e^{- I^{\textrm{\it on shell}}_E} = e^{-\b \F}, \label{se1:partitionfunction}
  }
where $g$ and $h$ denote the bulk and boundary values of the metric, $I_E$ is the Euclidean action, and $\b$ is the inverse Hawking temperature. The Euclidean action must be finite when evaluated on shell. It must also have a well-defined variational principle compatible with Dirichlet boundary conditions that fix the metric (and any additional fields) at the boundary. 

Higher spin theories are known for lacking an action principle from which their equations of motion may be derived~\cite{Fradkin:1987ks,Fradkin:1986qy,Vasiliev:1990en,Vasiliev:1992av}, the exception being higher spin theories in three dimensions~\cite{Blencowe:1988gj,Bergshoeff:1989ns,Campoleoni:2010zq}. Indeed, in parallel with three-dimensional gravity~\cite{Achucarro:1987vz,Witten:1988hc}, higher spin theories in AdS$_3$ admit a Chern-Simons formulation with gauge group $SL(N,R)\times SL(N,R)$ describing fields of spin 2, 3, $\dots$, N \cite{Campoleoni:2010zq}. However, when evaluated on shell, the Chern-Simons action of $SL(N,R)$ black holes generically vanishes and the covariant approach to the thermodynamics of higher spin black holes fails. The same problem is encountered in the Chern-Simons formulation of AdS$_3$ gravity and several ways to recover the thermodynamics of black holes were devised in~\cite{Banados:1993qp,Banados:1998ys} and generalized to higher spin theories in~\cite{Banados:2012ue,deBoer:2013gz}.

In this paper we show how to recover a finite, covariant, on-shell action for higher spin black holes. This is achieved by supplementing the Chern-Simons formulation of higher spin theories with boundary terms at the asymptotic boundary and, for black holes in Schwarzschild coordinates, with a boundary term at the horizon.\footnote{Although the term at the horizon may vanish in different coordinates, we expect the on-shell action of the higher spin theory to be independent of coordinates, in analogy with the spin-2 case considered below.} The former are a generalization of the boundary terms accompanying the Einstein-Hilbert action in the metric formulation of three-dimensional gravity~\cite{Gibbons:1976ue,Balasubramanian:1999re}. They guarantee both finiteness of the action and a well-defined variational principle that fixes the metric and higher spin fields at the boundary. We argue that these terms, otherwise non-linear in the Chern-Simons fields, become quadratic expressions of the latter in the Fefferman-Graham gauge.

The other boundary term is located at the horizon of the higher-spin black hole and is a generalization of the Gibbons-Hawking term for higher spin theories. This term arises when deriving the Chern-Simons formulation from the first order formulation of the theory. 
Although we do not treat the latter as a true boundary and we do not impose boundary conditions there, this boundary term does make a finite contribution to the action and is partly responsible for the free energy of higher spin black holes. Note that adding this term to the on shell action is equivalent to working with the first order formulation of the theory. Nevertheless we will stick to the Chern-Simons formulation since calculations are much simpler in the Chern-Simons language.

Focusing on the $SL(3,R)$ higher spin theory in the principal embedding, we will show that in the Fefferman-Graham gauge the on-shell action of higher spin black holes reads
  \eq{
  I^{\textrm{on shell}} = \frac{k_3}{4\pi} \int_{\H} \tr \( A \we \bar{A} \) - \dsint\( \mu \W + \bar{\mu} \overline{\W} \)  dt d\phi , \label{se1:onshell}
  }
where $k_3$ is inversely proportional to Newton's constant, $A$ and $\bar{A}$ are the $SL(3,R)$-valued Chern-Simons fields, $\mu$ and $\bar{\mu}$ are the sources for the higher spin charges $\W$ and $\overline{\W}$, and $\H$ and $\p \ss$ denote respectively the horizon and the asymptotic boundary.\footnote{The existence of a horizon for higher spin black holes is a non-trivial requirement since higher spin gauge transformations act non-trivially on the metric~\cite{Gutperle:2011kf,Castro:2011fm,Ammon:2011nk}. We assume that it is always possible, although technically challenging (see e.g.~\cite{Ammon:2011nk}), to find Schwarzschild-like coordinates where the horizon is manifest.} The asymptotic boundary term probes aspects of the induced theory of $\W$-gravity at the boundary (see e.g.~\cite{Schoutens:1991wm,Hull:1993kf} for reviews). In particular, general covariance of the latter implies that the higher spin charges do not obey the standard (chiral) Ward identities associated with the $\W_3 \times \W_3$ asymptotic symmetries of the theory.

Although the higher spin black hole of Ammon, Gutperle, Kraus, and Perlmutter (AGKP)~\cite{Gutperle:2011kf,Ammon:2011nk} is not in the Fefferman-Graham gauge, we find that eq.~\eqref{se1:onshell} leads to consistent thermodynamics in agreement with the results of~\cite{Perez:2013xi,Campoleoni:2012hp,Banados:2012ue,deBoer:2013gz} (see~\cite{Perez:2014pya} for additional references). However, in contrast to the higher spin black holes in the Fefferman-Graham gauge, the charges of the AGKP black hole do obey the chiral $\W_3$ Ward identities. This suggests that the AGKP boundary conditions lead to a different theory of induced $\W$-gravity at the boundary from that obtained from boundary conditions compatible with the Fefferman-Graham gauge.\footnote{The AGKP black hole can be put in the Fefferman-Graham gauge at the cost of violating the boundary conditions that lead to eq.~\eqref{se1:onshell}.}

The paper is organized as follows. In Section~\ref{se:sl2} we write down the boundary terms in the Chern-Simons formulation of three-dimensional gravity that guarantee a well-defined variational principle and a finite, non-vanishing, on-shell action. In Section~\ref{se:sln} we generalize these boundary terms to $SL(3,R)$ higher spin theories in the principal embedding and write down the covariant on-shell action of $SL(3,R)$ higher spin black holes. We compute the free energy and entropy of the AGKP higher spin black hole and interpret the asymptotic boundary terms in the on-shell action in the context of induced $\W$-gravity at the boundary. We end with our conclusions in Section~\ref{se:conclusions}. Our conventions are confined to Appendix~\ref{ap:conventions} while the generalization of our results to $SL(N,R)$ higher spin theories is considered in Appendix~\ref{ap:sln}.


\section{Warming up with $\mathbf{SL(2,R)}$}
\label{se:sl2}

In this section we introduce the asymptotic boundary terms in the Chern-Simons formulation of AdS$_3$ gravity in the Fefferman-Graham gauge that lead to a well-defined variational principle compatible with Dirichlet boundary conditions. For black holes in Schwarzschild-like coordinates where the horizon is manifest we also write down the boundary terms at the horizon that render the covariant, Euclidean, on-shell action finite and reproduce the free energy of the BTZ black hole. These boundary terms are generalized to the $SL(3,R)$ higher spin theory in the next section.

\subsection{Asymptotic boundary terms} 
\label{suse2:asymptotic}

Let us begin by discussing the asymptotic boundary terms in the Chern-Simons formulation of three-dimensional gravity with a negative cosmological constant. We would like to find boundary terms that lead to a well-defined variational principle compatible with Dirichlet boundary conditions for the metric. These boundary terms must also render the action free of divergences.\footnote{Note that the Chern-Simons action is finite on shell and its variation is finite as well. However, introducing boundary terms that fix the metric at the boundary spoils the finiteness of the action. Therefore additional boundary terms are necessary to regulate the action.} The boundary terms we are looking for are precisely those that, in the second order formulation of General Relativity, guarantee a well-defined variational principle and a finite action~\cite{Gibbons:1976ue,Balasubramanian:1999re}
  \eq{
  I_{GR} = \frac{k}{4\pi} \bigg \{ & \sint \absq{g} \(R + \frac{2}{\ell^2}\)d^3x + 2\dsint \absq{h} K\,d^2x - \frac{2}{\ell} \dsint \absq{h}\,d^2x \bigg \}, \label{se2:gr}
  }
where $k = 1/4G_N$, $G_N$ is Newton's constant, $\ell$ is the radius of AdS which we set to one, $h$ is the determinant of the induced metric at the boundary, and $K$ is the trace of the extrinsic curvature. 

As shown in refs.~\cite{Achucarro:1987vz,Witten:1988hc} three-dimensional gravity with a negative cosmological constant can be described by two Chern-Simons actions with opposite sign. Indeed, keeping track of boundary terms we have
  \eqsp{
  \frac{k}{4\pi}\sint \absq{g}\( R + 2 \) d^3x = & \,\,\,I_{CS}[A] - I_{CS}[\bar{A}] - \frac{k}{4\pi} \sint d[ \tr \( A \we \bar{A} \)] ,  \label{se2:eh2cs} 
  }
where $A$ and $\bar{A}$ are two $SL(2,R)$ gauge fields defined in terms of the dreibein $e^a$ and dual spin connection $w^a = \frac{1}{2} \e^{abc} w_{bc}$ by~\cite{Witten:1988hc}
  \eq{
  A^a = \( w^a + e^a \), \qquad \bar{A}^a  = \( w^a - e^a \), \label{se2:vielbeins}
  }
and $I_{CS}[A]$ is the Chern-Simons action given by
  \eq{
  I_{CS}[A] = \frac{k}{4\pi}\sint \tr \(A\we d A + \frac{2}{3} A\we A \we A \).
  }
The asymptotic boundary term in eq.~\eqref{se2:eh2cs} is proportional to the Gibbons-Hawking term, namely~\cite{Banados:1998ys}
  \eq{
 \dsint \tr (A\we \bar{A}) = 2\dsint \tr (e \we w) = \dsint \absq{h} K, \label{se2:gibbonshawking}
  }
where $\p \ss$ denotes the asymptotic boundary. Thus, the Chern-Simons formulation of three-dimensional gravity reads
  \eq{
  I[A,\bar{A}] = I_{CS}[A] - I_{CS}[\bar{A}] + I_b[A,\bar{A}], \label{se2:csaction}
  }
where, by construction, the boundary term $I_b[A,\bar{A}]$ leads to a finite action equipped with a well-defined variational principle that fixes the metric at the boundary. The boundary term is given by
  \eq{
   I_b[A,\bar{A}] = &\frac{k}{4\pi} \dsint \tr \(A\we \bar{A}\) - \frac{k}{8\pi} \dsint \Big | \textrm{det }\tr \lb (A-\bar{A})_{\mu} (A-\bar{A} )_{\nu} \rb \Big | ^{\frac{1}{2}} d^2x , \label{se2:csboundary}
  }
where Greek indices denote boundary coordinates.

The non-linear nature of the boundary term  $I_b[A,\bar{A}]$ is unappealing and makes the Chern-Simons formulation cumbersome. However, in the Fefferman-Graham gauge this term becomes quadratic in the Chern-Simons fields as we will soon find. Recall that in Fefferman-Graham coordinates the metric is given by~\cite{Fefferman:1985ok}
  \eq{
  d s^2 & = g_{MN} d x^{M}d x^{N} = \frac{d r^2}{r^2} + \( r^2 g^{(0)}_{\mu\nu} + g^{(2)}_{\mu\nu} + r^{-2} g^{(4)}_{\mu\nu} \) d x^{\mu} d x^{\nu}, \label{se2:fg}
  }
where uppercase Roman indices denote bulk spacetime coordinates and the boundary is located at $r \ra \infty$. In eq.~\eqref{se2:fg} $g^{(0)}_{\mu\nu}$ denotes the boundary metric while $g^{(2)}_{\mu\nu}$ encodes the charges of the solution, e.g.~its mass and angular momentum. In terms of the Chern-Simons fields the metric is given by
  \eq{
  g_{MN} = 2\, \tr \(e_M e_N\) = \frac{1}{2} \tr (A-\bar{A})_M (A-\bar{A})_N. \label{se2:fields}
  }
Thus, in the Chern-Simons formulation the Fefferman-Graham gauge corresponds to~\cite{Coussaert:1995zp,Banados:1998gg}
  \eqsp{
    A_r &= b^{-1} \p_r b, \qquad   A_{\mu} = b^{-1} a_{\mu}(x^{\pm}) b,  \qquad b = e^{\log(r) L_0}, \\
  \bar{A}_r &= \bar{b}^{-1} \p_r \bar{b}, \qquad  \bar{A}_{\mu} = \bar{b}^{-1} \bar{a}_{\mu}(x^{\pm}) \bar{b}, \qquad \bar{b} = b^{-1}, \label{se2:fgcs}
  }
supplemented by the following condition on the gauge fields~\cite{Banados:2002ey,Banados:2004nr}
  \eq{
  \tr \lb (A-\bar{A})_{\mu}\, L_0 \rb = 0. \label{se2:fgcs2}
  }
In eqs.~\eqref{se2:fgcs} and~\eqref{se2:fgcs2} $x^{\pm} = t \pm \phi$ denote boundary lightcone coordinates while $L_0$ is the zero mode of the $SL(2,R)$ algebra (see Appendix~\ref{ap:conventions} for our conventions). Note that it is always possible to enforce eq.~\eqref{se2:fgcs2} via residual gauge transformations compatible with eq.~\eqref{se2:fgcs}~\cite{Li:2015osa}.

In the Fefferman-Graham gauge the non-linear boundary term given in eq.~\eqref{se2:csboundary} becomes quadratic in the gauge fields
  \eq{
   I_b[A,\bar{A}] = \frac{k}{4\pi} \dsint \tr \( A\we \bar{A} \) - \frac{k}{4\pi}\dsint \tr \lb (A-\bar{A}) \we (A-\bar{A}) L_0 \rb.  \label{se2:csboundaryfinal}
  }
Unlike eq.~\eqref{se2:csboundary}, eq.~\eqref{se2:csboundaryfinal} can be easily generalized to higher spin theories. Note that since we have fixed the gauge, the Chern-Simons action is covariant despite the explicit appearance of $L_0$ in the boundary term. This is not surprising given the distinguished role that $L_0$ plays in the Fefferman-Graham gauge. 

The second term in eq.~\eqref{se2:csboundaryfinal} is one of a few generally-covariant choices that is quadratic in the fields. Any such term must depend on the combination $A -\bar{A} = 2 e$ since the alternative $A + \bar{A} = 2 w$ leads to an ill-defined variational principle in the first order formulation of the theory. Indeed, a different way to arrive at eq.~\eqref{se2:csboundaryfinal} is by demanding an action whose variation is finite and proportional to
  \eq{
  \d I[A,\bar{A}] \propto \dsint (\dots) \we \d (A - \bar{A}) \propto \dsint (\dots) \we \d e, 
  }
since such a variation is guaranteed to fix the metric at the boundary. Any boundary terms quadratic in $A-\bar{A}$  other than that appearing in eq.~\eqref{se2:csboundaryfinal} either introduce divergences in the action or vanish exactly in the Fefferman-Graham gauge. Nevertheless, it is possible to relax some of the conditions we have imposed on the action in interesting ways and obtain different but closely related boundary terms~\cite{Arcioni:2002vv,Apolo:2015fja}. 

One may wonder why have we derived the asymptotic boundary terms by appealing first to the second order formulation of three-dimensional gravity. The point is that eq.~\eqref{se2:csboundaryfinal} is valid \emph{only} in the Fefferman-Graham gauge. Away from the Fefferman-Graham gauge we should use the highly non-linear boundary term given in eq.~\eqref{se2:csboundary}. Any other choice of boundary terms, whether or not quadratic in the Chern-Simons fields, cannot be written in a generally-covariant form in the second order formulation of the theory. Since different boundary terms can give different finite contributions to the on-shell action (and still regulate all divergences) we need an additional principle to determine the appropriate boundary terms. This principle is general covariance of the gravitational theory the Chern-Simons formulation aims to describe.

Let us now consider the variation of the action. It is convenient to parametrize the gauge fields $a$ and $\bar{a}$ appearing in eq.~\eqref{se2:fgcs} as follows,
  \eq{
  a= \sum_a \l ^a L_a, \qquad \bar{a} = \sum_a \bar{\l}^a L_a, \qquad a = 0, \pm 1,
  }
where $L_a$ are the generators of $SL(2,R)$ in the highest-weight basis (see Appendix~\ref{ap:conventions}). In terms of these variables the boundary metric in eq.~\eqref{se2:fg} is given by
  \eq{
  g^{(0)}_{\mu\nu} = \l^{+1}_{(\mu}\, \bar{\l}^{-1}_{\nu)}, \label{se2:boundarymetric}
  }
where the indices are symmetrized with unit weight. We thus learn that in the Fefferman-Graham gauge highest-weight fields in the unbarred sector and lowest-weight fields in the barred sector determine the boundary data. Using this parametrization of the gauge fields, variation of the action yields
  \eq{
  \d I[A,\bar{A}] = \frac{k}{2\pi} \dsint \( \l^{-1} \we \d \l^{+1} - \bar{\l}^{+1} \we \d \bar{\l}^{-1} \), \label{se2:variation}
  }
where we have ignored the bulk term responsible for the equations of motion 
  \eq{0 = F = dA + A\we A , \qquad 0 = \bar{F} = d\bar{A} +\bar{A} \we \bar{A}. \label{se2:eom}
  }
Thus the boundary terms in eq.~\eqref{se2:csboundaryfinal} lead to a finite action with a well-defined variational principle that fixes the metric at the boundary. This should not be surprising since these terms were borrowed from the second order formulation of the theory. 

In particular, the most general solution compatible with Brown-Henneaux boundary conditions~\cite{Brown:1986nw},
 \eq{
  g_{\mu\nu} & = r^2 \eta_{\mu\nu} + \O(r^0), \qquad g_{rr} = r^{-2} + \O(r^{-4}), \qquad g_{r \mu} = \O(r^{-3}), \label{se2:bhbc}
  }
where $\eta_{\mu\nu}$ is the two-dimensional Minkowski metric, may be obtained by setting
  \eq{
  \l^{+1}_+ = 1, \qquad \bar{\l}^{-1}_- = -1, \qquad \l^{+1}_- = \bar{\l}^{-1}_+ = 0, \label{se2:bc}
  }
and solving the equations of motion~\eqref{se2:eom}. There are six components to the latter that allow us to solve for the six unknowns, the most general solution being~\cite{Banados:1998gg}
  \eq{
  a & = \( L_{+1} - \frac{2\pi}{k} \L \, L_{-1} \) dx^+, \qquad \bar{a} = -\( L_{-1} - \frac{2\pi}{k} \bar{\L} \, L_{+1} \) dx^-, \label{se2:ads}
  }
where $\L = \L(x^+)$ and $\bar{\L} = \bar{\L}(x^-)$ parametrize the space of solutions compatible with the boundary conditions.

The normalization of the subleading components of the unbarred and barred gauge fields is chosen so that the charges generating the asymptotic symmetries, described by two copies of the Virasoro algebra, are given by $\L$ and $\bar{\L}$, see e.g.~\cite{Banados:1998gg}. That these are the charges of the theory can also be seen by computing the holographic one-point functions of the dual stress-energy tensor. Indeed, using~\cite{Balasubramanian:1999re}
  \eq{
  \Vev{T_{\mu\nu}} = \frac{2}{\absq{g^{(0)}}} \frac{\d I}{\d g^{(0)\mu\nu}}, \label{se2:tmunu}
  }
and eqs.~\eqref{se2:boundarymetric} and~\eqref{se2:variation} we find
  \eq{
  \Vev{T_{++}} &= \frac{1}{\bar{\l}^{-1}_-} \frac{\d I}{\d \l^{+1}_-} = \L, \qquad \Vev{T_{--}} = \frac{1}{\l^{+1}_+} \frac{\d I}{\d \bar{\l}^{-1}_+} = \bar{\L}, \label{se2:onepoint}
  }
while the $\Vev{T_{+-}}$ component vanishes as expected from the conformal symmetry of the dual theory.


\subsection{On-shell action and black hole thermodynamics} 
\label{suse2:onshellaction}

Let us now reconsider the thermodynamics of the BTZ black hole in the Chern-Simons formulation. When deriving the Chern-Simons formulation of three-dimensional gravity we dropped a total derivative contribution corresponding to the lower limit of integration of the bulk integral in eq.~\eqref{se2:eh2cs}. In Euclidean signature the locus of this contribution is the origin of coordinates and depending on the coordinate system used this term may or may not contribute to the on-shell action.\footnote{Note that the on-shell action with \emph{all} total derivative terms included is independent of coordinates, i.e.~it corresponds to the on-shell Einstein-Hilbert action.} In this paper we will consider Schwarzschild-like coordinates where the location of the horizon in Lorentzian signature corresponds to the origin of coordinates in Euclidean signature. Thus, in this coordinate system the on-shell action receives a non-trivial contribution from a boundary term at the horizon. Since we do not treat the horizon as a real boundary in the Lorentzian theory, i.e.~we do not impose boundary conditions there, we are justified in ignoring this boundary term in the off-shell action. Nevertheless, this term does give a non-vanishing contribution to the Euclidean on-shell action and is partly responsible for the free energy of the BTZ black hole. From eqs.~\eqref{se2:eh2cs} and~\eqref{se2:gibbonshawking} we see that this boundary term is nothing but the Gibbons-Hawking term at the horizon. 

The covariant, on-shell (os) action of three-dimensional gravity in the Chern-Simons formulation is thus given by
  \eq{
  I^{os}[A,\bar{A}] = & \,\,\,I^{os}_{CS}[A] - I^{os}_{CS}[\bar{A}] + I^{os}_b[A,\bar{A}] + \frac{k}{4\pi} \int_{\H} \tr \( A \we \bar{A} \), \label{se2:onshell}
  }
where $\H$ denotes the location of the horizon. Note that by keeping all the boundary terms that arise in the derivation of the Chern-Simons formulation of three-dimensional gravity we are effectively working with the first order formulation of the theory written in different variables. Thus, it is not surprising that we can reproduce the thermodynamics of the BTZ black hole. Nevertheless one advantage of the Chern-Simons formulation is that the Chern-Simons actions $I_{CS}[A]$ and $I_{CS}[\bar{A}]$ vanish in the gauge~\eqref{se2:fgcs} so that the covariant, on-shell action is given entirely by the boundary terms
  \eq{
  I^{os}[A,\bar{A}] = \frac{k}{4\pi} \int_{\H} \tr \( A \we \bar{A} \) + I^{os}_b[A,\bar{A}]. \label{se2:onshell2}
  }
This will prove to be especially convenient for the higher spin black hole whose connections depend non-trivially on the radial coordinate.

For completeness let us conclude this section by obtaining the thermodynamic properties of the non-rotating BTZ black hole in the canonical ensemble. The latter is described by the solution given in eqs.~\eqref{se2:ads} with $\L = \bar{\L} \ge 0$ and its metric is given by~\cite{Banados:1998gg}
  \eq{
  d s^2  =&\,\,\, \frac{d r^2}{r^2} - \frac{1}{2} \( r^2 + \frac{1}{r^2} \frac{4\pi^2}{k^2} \L^2 \) (dt^2-d\phi^2) - \frac{4\pi}{k} \L\, (dt^2+d\phi^2).   \label{se2:btz}
  }
In the semi-classical approximation the free energy $\F$ of the BTZ black hole can be obtained from the Euclidean partition function
  \eq{
  Z = \int \D [A, \bar{A}] \, e^{- I_E[A,\bar{A}] } \sim e^{- I^{os}_E[A,\bar{A}]} = e^{-\b \F}, \label{se2:partitionfunction}
  }
where $I_E^{os}$ is the covariant, Euclidean, on-shell action and $\b$ is the inverse Hawking temperature. The latter is determined from the identification $t_E \sim t_E + \b$ that guarantees a smooth horizon free of conical singularities in Euclidean signature. For the metric given in eq.~\eqref{se2:btz}  the temperature is given by
  \eq{
  \b = \frac{2\pi}{r_{\H}} \sqrt{\frac{2}{- \p_r^2 g_{tt}}} = \sqrt{\frac{\pi k}{2\L}},
  }
where $r_\H$ is the location of the horizon,
  \eq{
  r_{\H} = \sqrt{\frac{2\pi}{k}\L}. \label{se2:horizon}
  }
Alternatively, the temperature of the BTZ black hole, which has the topology of a solid torus in Euclidean signature, can be determined by demanding trivial holonomies for the gauge fields~\eqref{se2:ads} around the contractible (time) cycle (see e.g.~\cite{Ammon:2012wc}).

In the Fefferman-Graham gauge the Chern-Simons actions $I_{CS}[A]$, $I_{CS}[\bar{A}]$, and the boundary term $I_b[A,\bar{A}]$ all vanish on shell. Thus, the only contribution to the free energy comes from the boundary term at the horizon,
  \eq{
  \F = -\frac{1}{\b} \frac{k}{4\pi} \int_{\H} \tr \( A \we \bar{A} \) =  -\frac{2\pi^2 k}{\b^2}, \label{se2:freeenergy}
  }
where the relative minus sign with respect to eq.~\eqref{se2:onshell} comes from the Euclidean continuation, i.e.~$t \ra -i t_E$. From the free energy one readily obtains the entropy
  \eq{
  \cS &= \b^2 \p_{\b} \F = 4 \pi \sqrt{2\pi k \L},
  }
which agrees with the area law of black hole entropy, and the charge conjugate to the inverse temperature,
  \eq{
  \cE &= \p_{\b} (\b\F) = 4\pi \L,
  }
 which agrees with the $T_{tt}$ component of the stress-energy tensor~\eqref{se2:onepoint} up to normalization. As expected, from these equations it follows that
  \eq{
  \F = \cE - \cS/\b.
  }
  %
 

\section{SL(3,R) higher spin theory} 
\label{se:sln}

In this section we consider the generalization of the boundary terms considered in the previous section to higher spin theories. These boundary terms lead to a finite action with a well-defined variational principle that fixes the metric and higher spin fields at the boundary. We focus on the $SL(3,R)$ higher spin theory in the principal embedding. Comments on the generalization to $SL(N,R)$ higher spin theories are given in appendix~\ref{ap:sln}. 


\subsection{Asymptotic boundary terms in the Fefferman-Graham gauge} 
\label{suse3:asymptotic}

Let us begin by discussing the asymptotic boundary terms of the $SL(3,R)$ higher spin theory. In parallel with three-dimensional gravity with a negative cosmological constant, this theory may be formulated in terms of two $SL(3,R)$ Chern-Simons actions with opposite sign~\cite{Blencowe:1988gj,Bergshoeff:1989ns,Campoleoni:2010zq},
  \eq{
  I_{3}[A,\bar{A}] = I_{CS}[A] - I_{CS}[\bar{A}] + I_{b(3)}[A,\bar{A}], \label{se3:sl3action}
  }
where $I_{b(3)}[A,\bar{A}]$ is to be determined and $I_{CS}$ carries a different normalization from its $SL(2,R)$ counterpart (see Appendix~\ref{ap:conventions} for our conventions),
  \eq{
  I_{CS}[A] = \frac{k_3}{4\pi}\sint \tr \(A\we d A + \frac{2}{3} A\we A \we A \),
  }
where $k_3 = k/4 = 1/16G_N$. The $SL(3,R)$ gauge fields are defined in terms of a generalized dreibein and (dual) spin connection
  \eq{
  A^a &= \( \tilw^a + \tile^a \), \qquad \bar{A}^a = \( \tilw^a - \tile^a \), \qquad a= 1,\dots, 8. \label{se3:vielbeins}
  }
In particular, the theory also admits a first order formulation similar to General Relativity but in terms of these objects,
  \eq{
  \tilde{I} = \frac{k}{4\pi} \sint \tr \( \tile\we d \tilw + \tile\we \tilw\we \tilw  + \frac{1}{3} \tile \we \tile \we \tile  \). \label{se3:firstorder}
  }

The spectrum of higher spin theories depends on how the gravitational $SL(2,R)$ sector is embedded in $SL(N,R)$. In this paper we will focus on the principal embedding where the actions~\eqref{se3:sl3action} and~\eqref{se3:firstorder} describe the non-linear interactions of spin-2 and spin-3 gauge fields. The latter may be recovered from the Chern-Simons fields or the generalized dreibeins via
  \eq{
  g_{MN} = \frac{1}{2} \tr (\tile_M \tile_N), \qquad \psi_{MNP} = \frac{1}{3} \tr \lb \tile_{(M} \tile_N \tile_{P)} \rb. \label{se3:fields}
  }
These expressions reveal an interesting feature of higher spin theories: both diffeomorphisms and spin-3 gauge transformations act non-trivially on the metric.  

The similarities between eqs.~\eqref{se3:vielbeins} and~\eqref{se3:fields} and their $SL(2,R)$ counterparts, eqs.~\eqref{se2:vielbeins} and~\eqref{se2:fields}, suggest that the bulk contribution to the action~\eqref{se3:sl3action} does not have a variational principle compatible with Dirichlet boundary conditions for the metric and spin-3 field. This is most easily seen in the first order formulation of the theory whose variation yields
  \eq{
  \d \tilde{I} = -\frac{k}{4\pi} \dsint \tile \we \d \tilw, \label{se3:varfirstorder}
  }
where we have ignored the bulk term that yields the Chern-Simons equations of motion $F = \bar{F} = 0$.
  
The boundary term that yields the desired variation of the action, i.e.~one proportional to $\dsint \(\dots\) \we \d (A-\bar{A})$, is given by
  \eq{
   I_{b(3)}[A,\bar{A}] = \frac{k_3}{4\pi} \dsint \tr \(A\we \bar{A}\) + B_{3}[A-\bar{A}], \label{se3:csboundary}
  }
where $\dsint \tr \(A \we \bar{A}\) = 2\dsint \tr \( \tile\we \tilw \)$ is the generalization of the Gibbons-Hawking term appropriate for higher spin theories, cf.~eq.~\eqref{se2:gibbonshawking}. The first boundary term can be obtained by deriving the Chern-Simons action from the first order action given in eq.~\eqref{se3:varfirstorder} and introduces divergences to the otherwise finite action. The counterterm $B_{3}[A-\bar{A}]$ cancels these divergences and leads to a finite on-shell action and finite one-point functions for the dual stress-energy tensor and spin-3 current. In analogy with pure three-dimensional gravity we expect the counterterm $B_3[A,\bar{A}]$ to be non-linear in the Chern-Simons fields. This is a consequence of general covariance of the second order formulation of higher spin theories. It is also not unreasonable to expect that, in the Fefferman-Graham gauge, this term becomes quadratic in the Chern-Simons fields.
  
The Fefferman-Graham gauge for the $SL(3,R)$ connections $A$ and $\bar{A}$ is given by eq.~\eqref{se2:fgcs}, where $L_0$ is now the zero mode of the $SL(2,R)$ subalgebra, supplemented by the following conditions~\cite{Campoleoni:2010zq}
  \eq{
  \tr \lb (A-\bar{A})_{\mu}\, L_0 \rb = 0, \qquad \tr \lb (A-\bar{A})_{\mu}\, W_0 \rb = 0, \label{se3:fgcs3}
  }
where $W_0$ is one of the generators associated with the spin-3 part of the algebra. While the first condition guarantees $g_{\mu r} = 0$, the second condition implies $\psi_{\mu rr} = 0$. Note, however, that these conditions do not fully fix the gauge. One must also impose~\cite{Campoleoni:2010zq}
  \eq{
  \tr \( A_{+}\, W_{+1} \) = 0, \qquad \tr \( \bar{A}_{-} \, W_{-1} \) = 0. \label{se3:fgcsextra}
  }

It is important to note that the Fefferman-Graham gauge is compatible with a large set of yet to be determined boundary conditions on the metric and spin-3 field. However, if we move away from eq.~\eqref{se3:fgcs3} we automatically impose boundary conditions for the $g_{\mu r}$ and $\psi_{\mu r r}$ components of these fields. Indeed, relaxing the first and second conditions in eq.~\eqref{se3:fgcs3} we find, respectively,
  \eq{
  g_{\mu r} \sim \frac{1}{r}, \qquad \psi_{\mu r r} \sim \frac{1}{r^2}. \label{se3:nonfgbc}
  }
In particular note that $g_{\mu r}$ is not compatible with Brown-Henneaux boundary conditions~\eqref{se2:bhbc}. If we relax eq.~\eqref{se3:fgcs3} the gauge symmetry may be used to fix instead
  \eq{
  \tr \( A_{+}\, L_0 \) = \tr \( \bar{A}_{-}\, L_0 \) = 0, \qquad \tr \( A_{+}\, W_0 \) = \tr \( \bar{A}_{-}\, W_0 \) = 0.
  }

Now that we are in the Fefferman-Graham gauge we may consider turning off the higher spin fields. Then $I_{b(3)}[A,\bar{A}]$ must reduce to the boundary term $I_b[A,\bar{A}]$ of the $SL(2,R)$ theory. This means that up to contributions from the higher spin generators $W_a$, the asymptotic boundary terms of the higher and lower spin theories must match. Any additional contributions are then fixed by requiring cancellation of any residual divergences in the action. We thus obtain,
  \eqsp{
   I_{b(3)}[A,\bar{A}] = &\:\frac{k_3}{4\pi}\; \dsint \tr \( A\we \bar{A} \) - \frac{k_3}{4\pi}\dsint \tr \lb (A-\bar{A}) \we (A-\bar{A}) L_0 \rb \\
   +& \frac{k_3}{64\pi} \dsint \tr \lb(A-\bar{A}) W_{+2} \rb\we \tr \lb(A-\bar{A}) W_{-2} \rb.\label{se3:csboundaryfinal}
  }

The boundary terms in eq.~\eqref{se3:csboundaryfinal} could have been obtained by demanding a finite action whose variation fixes the spin-2 and spin-3 fields at the boundary without any reference to the $SL(2,R)$ case. One may then be tempted to conclude that the second and third terms in eq.~\eqref{se3:csboundaryfinal} are appropriate for other gauges as well since, for example, relaxing eq.~\eqref{se3:fgcs3} does not introduce new divergences in the action. This is where comparison to the $SL(2,R)$ case is important: if we relax the analogous condition for $SL(2,R)$, namely eq.~\eqref{se2:fgcs2}, then the appropriate boundary term is highly non-linear in the Chern-Simons fields. Although both counter-terms in eqs.~\eqref{se2:csboundary} and~\eqref{se2:csboundaryfinal} regularize the action, their finite contributions to the on-shell action are different. We expect the appropriate boundary term to be the one that can be expressed in a generally-covariant form in the metric formulation of the theory.

Similar statements hold for higher spin theories. In particular we expect the boundary terms in the second order formulation of the theory to be highly non-linear expressions of the Chern-Simons variables. We also expect these boundary terms to be uniquely determined by general covariance of the metric formulation of the theory. In analogy to the $SL(2,R)$ case, we conjecture that in the Fefferman-Graham gauge these boundary terms become quadratic in the Chern-Simons fields and are given by eq.~\eqref{se3:csboundaryfinal}. Indeed, while it is clear that away from the Fefferman-Graham gauge eq.~\eqref{se3:csboundaryfinal} becomes highly non-linear in the gauge fields, we do not prove that the converse is true, namely that eq.~\eqref{se3:csboundaryfinal} is the \emph{Fefferman-Graham limit} of a term covariant in the metric and spin-3 field.

Let us now show that eq.~\eqref{se3:csboundaryfinal} leads to a finite action with a well-defined variational principle that fixes the metric and higher spin fields at the boundary. It is once again convenient to parametrize the fields $a$ and $\bar{a}$ appearing in eq.~\eqref{se2:fgcs} as follows
  \eq{
  a &= \sum_a \l^a L_a + \sum_b \om^b W_b, \qquad \bar{a} = \sum_a \bar{\l}^a L_a + \sum_b \bar{\om}^b W_b,
  }
where $L_a$ ($a = 0,\pm 1$) and $W_a$ ($a=0,\pm 1,\pm 2$) are the generators of $SL(3,R)$ in the highest-weight basis. In the Fefferman-Graham gauge the boundary components of the metric read~\cite{Li:2015osa}
  \eq{
  g_{\mu\nu} = r^2 \sum_{n=-1}^3 r^{-2n} g^{(2n)}_{\mu\nu}, \label{se3:metric}
  }
where the leading components are given by
  \eq{
  g_{\mu\nu}^{(-2)} &= -4\, \om^{+2}_{(\mu}\, \bar{\om}^{-2}_{\nu)}, \qquad g_{\mu\nu}^{(0)} = \l^{+1}_{(\mu} \, \bar{\l}^{-1}_{\nu)} + \om^{+1}_{(\mu}\, \bar{\om}^{-1}_{\nu)}.
  }
On the other hand the boundary components of the spin-3 field are given by~\cite{Li:2015osa}
  \eq{
  \psi_{\mu\nu\a} = r^4 \sum_{n=0}^{4} r^{-2n} \psi^{(2n)}_{\mu\nu\a},
  }
whose leading component reads
  \eq{
  \psi_{\mu\nu\a}^{(0)} &=  \Big [ \om^{+1}_{(\mu}\,\om^{+1}_{\phantom{(}\nu} - \l^{+1}_{(\mu}\,\l^{+1}_{\phantom{(}\nu} \Big]\,\bar{\om}^{-2}_{\a)} - \Big [\bar{\om}^{-1}_{(\mu}\,\bar{\om}^{-1}_{\phantom{(}\nu} - \bar{\l}^{-1}_{(\mu}\,\bar{\l}^{-1}_{\phantom{(}\nu} \Big] \,\om^{+2}_{\a)}.
  }
Thus, the boundary values of the metric and spin-3 fields are determined by the highest and lowest-weight components of the unbarred and barred gauge fields, respectively. This is a generic feature of $SL(N,R)$ higher spin theories in the Fefferman-Graham gauge. Unlike the $SL(2,R)$ theory, however, not all of these components represent true boundary data, i.e.~not all of these fields are fixed by the action principle.

In terms of these variables variation of the action~\eqref{se3:sl3action} yields
  \eqsp{
  \d I_3[A,\bar{A}] = \frac{k}{2\pi} \dsint \Big\{ &\l^{-1} \we \d \l^{+1} - \bar{\l}^{+1} \we \d \bar{\l}^{-1} + \om^{-1} \we \d\om^{+1} -\bar{\om}^{+1}\we \d \bar{\om}^{-1}   \\
   &-4\, \( \om^{-2}\we \d \om^{+2} - \bar{\om}^{+2}\we \d\bar{\om}^{-2}\) \Big\}, \label{se3:variation}
  }
where we have ignored the bulk term responsible for the equations of motion. Thus the boundary term~\eqref{se3:csboundaryfinal} performs as advertised: it yields a finite action with a well-defined variational principle that fixes the metric and spin-3 field at the boundary. Note that we have written eq.~\eqref{se3:variation} in a convenient way without yet imposing the additional gauge fixing given in eq.~\eqref{se3:fgcs3}. The latter equation implies that $\om^{+1}_-$ and $\bar{\om}^{-1}_+$ are not fixed by the boundary conditions but rather by the equations of motion.

The most general solution compatible with AdS boundary conditions~\cite{Campoleoni:2010zq} may be obtained by fixing the following boundary data,
  \eq{
  \l^{+1}_+ = 1, \qquad \bar{\l}^{-1}_- = - 1, \qquad \l^{+}_- = \bar{\l}^{-}_+ = \om^{+1}_+ = \bar{\om}^{-1}_- = \om^{+2}_{\mu} = \bar{\om}^{-2}_{\mu} = 0. \label{se3:bc1}
  }
The sixteen components of the equations of motion then allow us to solve for the sixteen unknowns. The solution is most easily expressed as~\cite{Campoleoni:2010zq},
  \eq{
  a & = \( L_{+1} - \frac{2\pi}{k} \L \, L_{-1} - \frac{\pi}{2k} \W\, W_{-2} \) dx^+, \label{se3:ads1} \\
  \bar{a} & = -\( L_{-1} - \frac{2\pi}{k} \bar{\L} \, L_{+1} -\frac{\pi}{2k} \overline{\W} \,W_{+2} \) dx^-, \label{se3:ads2}
  }
where 
  \eq{
  \p_- \L = \p_- \W = 0, \qquad \p_+ \bar{\L} = \p_+ \overline{\W} = 0.
  }
The functions $\L$, $\bar{\L}$, $\W$, and $\overline{\W}$ generate the asymptotic symmetries of the theory which are described by two copies of the $\W_3$ algebra~\cite{Campoleoni:2010zq}.

As in the $SL(2,R)$ case we can use eq.~\eqref{se3:variation} to compute the holographic one-point functions of the dual stress-energy tensor and spin-3 current. The former is given by eq.~\eqref{se2:tmunu} so that 
  \eq{
  \Vev{T_{++}} &= \frac{1}{\bar{\l}^{-1}_-} \frac{\d I_3}{\d \l^{+1}_-} = \L, \qquad \Vev{T_{--}} = \frac{1}{\l^{+1}_+} \frac{\d I_3}{\d \bar{\l}^{-1}_+} = \bar{\L}, \label{se3:onepointtmunu1}
  }
while the latter is naturally defined by
  \eq{
  \Vev{W_{\mu\nu\a}} = -\frac{4}{\absq{g^{(0)}}} \frac{\d I_3}{\d \psi^{(0)\mu\nu\a}}, \label{se3:spin3current}
  }
whose only non-vanishing components are given by
  \eq{
  \Vev{W_{+++}} &= \frac{1}{\bar{\l}^{-1}_- \bar{\l}^{-1}_-} \frac{\d I_3}{\d \om^{+2}_-} = \W, \qquad \Vev{W_{---}} = -\frac{1}{\l^{+1}_+ \l^{+1}_+} \frac{\d I_3}{\d \bar{\om}^{-2}_+} = \overline{\W}. \label{se3:onepointw1}
  }
Not surprisingly these quantities agree with the charges generating the asymptotic symmetries of the theory. Note that the dual stress-tensor in eq.~\eqref{se2:tmunu} and the dual spin-3 current in eq.~\eqref{se3:spin3current} are defined with respect to $g^{(0)}_{\mu\nu}$, which is the metric the dual theory couples to, instead of $g^{(-2)}_{\mu\nu}$, which is the leading component of the metric in eq.~\eqref{se3:metric}.
   

\subsection{The covariant action of higher spin black holes} 
\label{suse3:onshellaction}

Let us assume that higher spin black holes can be expressed in Schwarzschild-like coordinates, i.e.~in a coordinate system featuring a horizon which corresponds to the origin of coordinates in Euclidean signature. Then, the on-shell action of higher spin black holes receives a contribution from the generalized Gibbons-Hawking term at the horizon,
  \eq{
  I^{os}_3[A,\bar{A}] = & \,\,\,I^{os}_{CS}[A] - I^{os}_{CS}[\bar{A}] + I^{os}_{b(3)}[A,\bar{A}] + \frac{k_3}{4\pi} \int_{\H} \tr \( A \we \bar{A} \). \label{se3:onshell}
  }
In analogy to the $SL(2,R)$ case this term arises when deriving the Chern-Simons formulation of the higher spin theory from its first order formulation~\eqref{se3:firstorder}, i.e.~it corresponds to a total derivative term (cf.~eq.~\eqref{se2:eh2cs}). The fact that it can be interpreted as a boundary term at the horizon is a feature of our coordinate system. Nevertheless, we expect the on-shell action to be independent of the choice of coordinates as long as all the total derivative contributions that result from the derivation of the Chern-Simons action are properly taken into account. 

A generic feature of the Fefferman-Graham gauge, and in particular of eq.~\eqref{se2:fgcs}, is that the Chern-Simons action $I_{CS}[A]$ vanishes on-shell. Thus, although the generalized Gibbons-Hawking term at the horizon does not contribute to the variation of the action, i.e.~it is not part of the off-shell action, it makes a crucial contribution to the on-shell action. As in the $SL(2,R)$ theory we find that the on-shell action is given entirely by the boundary terms,
  \eq{
  I^{os}_3[A,\bar{A}] = \frac{k_3}{4\pi} \int_{\H} \tr \( A \we \bar{A} \) +  I^{os}_{b(3)}[A,\bar{A}] . \label{se3:onshell2}
  }

Let us now determine the on-shell value of the asymptotic boundary term. In order to do this we must turn on sources for the higher spin charges $\W$ and $\overline{\W}$. One motivation behind this is that eqs.~\eqref{se3:ads1} and~\eqref{se3:ads2} do not admit black hole solutions with higher spin charges. Indeed, the latter should be accompanied by conjugate chemical potentials which are naturally associated with sources for the higher spin fields~\cite{Gutperle:2011kf}. As seen from eq.~\eqref{se3:onepointw1} these sources correspond to non-vanishing values of the $\om^{+2}_{-}$ and $\bar{\om}^{-2}_{+}$ fields. Therefore let us consider a slight modification of the boundary conditions given in eq.~\eqref{se3:bc1} where $\om^{+2}_{-}$ and $\bar{\om}^{-2}_{+}$ do not vanish (for alternative boundary conditions see~\cite{Gutperle:2011kf,Castro:2011fm,Bunster:2014mua})
  \begin{gather}
  \l^{+1}_+ = 1, \qquad \bar{\l}^{-1}_- = - 1, \qquad \om^{+2}_- = \mu, \qquad \bar{\om}^{-2}_{+} = -\bar{\mu}, \label{se3:sourcebc1} \\
   \l^{+}_- = \bar{\l}^{-}_+ =  \om^{+1}_+ = \bar{\om}^{-1}_- =\om^{+2}_+ = \bar{\om}^{-2}_- = 0. \label{se3:sourcebc2}
  \end{gather}

The most general solution obeying these boundary conditions cannot be expressed algebraically in terms of the sources and its derivatives~\cite{Li:2015osa}. We may therefore consider constant solutions to the equations of motion which encompass higher spin black holes. These solutions are given by
  \eq{
  &a = \( L_{+1} - \frac{2\pi}{k} \L \,L_{-1} + \frac{4\pi}{k} \bar{\mu} \bar{\L} \,W_0- \frac{\pi}{2k}\W \,W_{-2} \)\, dx^+ \notag \\
   &+ \mu \lb W_{+2} + \frac{4\pi}{k} \W\,L_{-1} - \frac{4\pi}{k} \L\,W_0 + \frac{4\pi^2}{k^2}\( \L^2 - 2\bar{\mu}\bar{\L}\W \)W_{-2}  \rb dx^- \label{se3:sources1} \\
  &\bar{a} = -\( L_{-1} - \frac{2\pi}{k} \bar{\L} \,L_{+1} + \frac{4\pi}{k} \mu\L \, W_0 - \frac{\pi}{2k}\overline{\W} \, W_{+2} \) dx^- \notag \\
  &-\bar{\mu} \lb W_{-2} + \frac{4\pi}{k} \overline{\W}\,L_{+1} - \frac{4\pi}{k} \bar{\L}\,W_0 + \frac{4\pi^2}{k^2}\( \bar{\L}^2 - 2 \mu\L\overline{\W} \) W_{+2} \rb dx^+. \label{se3:sources2}
  }
Note that if these boundary conditions admit non-trivial asymptotic symmetries, the charges generating these symmetries are expected to be non-linear expressions of the charges of the undeformed theory~\cite{Perez:2012cf,Henneaux:2013dra,Compere:2013gja,Compere:2013nba}. 

It is also important to note that the charges corresponding to these boundary conditions do not obey the standard $\W_3$ Ward identities~\cite{Li:2015osa}. The latter may be obtained from the equations of motion by choosing chiral boundary conditions where one of the higher spin sources is turned off. For example for the unbarred sector we have
  \eq{
  \l^{+1}_+ = 1, \qquad \bar{\l}^{-1}_- = - 1, \qquad \om^{+2}_- = \mu, \label{se3:chiralsourcebc}
  }
with all other boundary data set to zero. The equations of motion then yield the $\W_3$ Ward identities for the spin-2 and spin-3 charges~\cite{Li:2015osa},
  \eq{
  \p_- \L =& -\( 3\,\p_+ \mu + 2\,\mu \,\p_+ \) \W, \label{se3:ward1} \\
  \p_- \W =& -\frac{k}{12\pi} \p_-^5 \mu + \(\frac{2}{3}\mu\,\p_+^3 + 3 \p_+\mu\,\p_+^2 + 5 \p_+^2 \mu\,\p_+ + \frac{10}{3} \p_+^3 \mu \) \L \notag \\
  & + \frac{32\pi}{3k} \(\mu\,\p_+ + 2\p_+\mu\)\L^2. \label{se3:ward2}
  }
On the other hand the charges corresponding to the boundary conditions given in eqs.~\eqref{se3:sourcebc1} and~\eqref{se3:sourcebc2} obey generalized Ward identities given by eqs.~\eqref{se3:ward1} and~\eqref{se3:ward2} where $\L$, $\W$, $\mu$ receive non-linear corrections in barred \emph{and} unbarred quantities~\cite{Li:2015osa}. 

Without sources for the higher-spin fields the boundary term~\eqref{se3:csboundaryfinal} vanishes on shell and the action is given by the generalized Gibbons-Hawking term at the horizon. On the other hand, once sources for the higher spin charges are turned on the covariant on-shell action becomes
  \eq{
  I^{os}_{3}[A,\bar{A}] = \frac{k_3}{4\pi} \int_{\H} \tr \( A \we \bar{A} \) - \dsint \(  \mu \W + \bar{\mu} \overline{\W} \) dt d\phi. \label{se3:onshellfinal}
  }
Note that all boundary terms were necessary to obtain this result: while only the last term in eq.~\eqref{se3:csboundaryfinal} gives a finite contribution to eq.~\eqref{se3:onshellfinal}, the other two boundary terms are required to remove $\O(r^2)$ and $\O(r^4)$ divergences.

The asymptotic boundary term takes the form of the (irrelevant) perturbation we would add to the dual conformal field theory once sources for the higher spin charges are turned on, except that the latter do not obey the $\W_3$ Ward identities. This may be understood as a consequence of the general covariance of the induced theory of $\W$-gravity at the boundary. In the context of the AdS/CFT correspondence~\cite{Maldacena:1997re,Gubser:1998bc,Witten:1998qj}, the on-shell action is sensitive to the Weyl and $\W$-Weyl anomalies of the dual conformal field theory~\cite{Henningson:1998gx,Skenderis:1999nb,Li:2015osa}. These anomalies play an important role in determining the induced action of two-dimensional gravity at the boundary~\cite{Polyakov:1987zb}. Indeed, for pure three-dimensional gravity the non-local Polyakov action may be derived directly by integrating the variation of the action given in eq.~\eqref{se2:variation}. 

It is not unreasonable to expect that a similar approach yields the induced action of $\W$-gravity in higher spin theories (see e.g.~\cite{Schoutens:1991wm,Hull:1993kf} for reviews and refs.~\cite{Poojary:2014ifa} and~\cite{Li:2015osa} for recent work in the context of higher spin theories). For chiral boundary conditions~\eqref{se3:chiralsourcebc}, solving the Ward identities given in eqs.~\eqref{se3:ward1} and~\eqref{se3:ward2} should yield the action of induced $\W$-gravity in the chiral gauge\footnote{Note that in contrast to ref.~\cite{Ooguri:1991by} we have turned off the source for the $T_{--}$ component of the stress-energy tensor.}~\cite{Ooguri:1991by},
  \eq{
  \d I_3[A,\bar{A}] = \dsint \W(\mu) \d \mu \,d^2x = \d \I_{\mu}, \label{se3:chiralwgravity}
  }
where $d^2x = dx^+ dx^- = 2 dt d\phi$ and we have set $\om^{-2}_+ = - \frac{\pi}{2k} \W$. In eq.~\eqref{se3:chiralwgravity} $\W(\mu)$ is a non-local function of $\mu$, namely the solution to the Ward identities, and $\I_{\mu}$ denotes the induced action of chiral $\W$-gravity. The latter is expected to be a covariant expression of boundary quantities, i.e.~the boundary metric $h_{\mu\nu}$ and higher spin source $\vp_{\mu\nu\a}$.\footnote{For the chiral boundary conditions in eq.~\eqref{se3:chiralsourcebc} we have $h_{\mu\nu} = \eta_{\mu\nu}$, where $\eta_{\mu\nu}$ is the Minkowski metric, and $\vp_{+++} = \mu$ while all other components vanish.} On the other hand, for the non-chiral boundary conditions given in eqs.~\eqref{se3:sourcebc1} and~\eqref{se3:sourcebc2} we expect to find 
  \eq{
  \d I_3[A,\bar{A}] = \dsint  \( \W\, \d \mu + \overline{\W}\, \d \bar{\mu} \) d^2x \ne \d \I_{\mu} + \d \I_{\bar{\mu}}, \label{se3:imtired}
  }
where we have used $\om^{-2}_+ = - \frac{\pi}{2k} \W$ and $\bar{\om}^{+2}_- = \frac{\pi}{2k} \overline{\W}$ in eq.~\eqref{se3:variation}. The reason for the inequality in eq.~\eqref{se3:imtired} is that we expect the non-chiral action of $\W$-gravity to be a non-local and covariant expression of the higher spin sources. Thus the higher spin charges cannot obey the standard $\W_3$ Ward identities in agreement with the results of~\cite{Li:2015osa}. In lieu of the induced action of $\W$-gravity, where one could explicitly check these statements, we have the on-shell value of the asymptotic boundary term~\eqref{se3:csboundaryfinal} (evaluated for constant sources and charges).

We conclude this section with a few comments. First, note that it was not necessary to determine the charges of the theory in order to obtain the on-shell action and consequently the thermodynamics of higher spin black holes. Thus, an advantage of the covariant approach is that, once the solution is known, its thermodynamics follow effortlessly. Furthermore, after going to Euclidean signature with $t \ra - it_E$ the asymptotic boundary term becomes,
  \eq{
  -i I_{b(3)}^{os} = \a \W + \bar{\a} \overline{\W},
  }
where the thermodynamic sources $\a$ and $\bar{\a}$ for the higher spin charges are given in terms of the inverse Hawking temperature $\b$ by
  \eq{
  \a = 2\pi \b \mu, \qquad \bar{\a} = 2\pi\b \bar{\mu},
  }
in agreement with the results of~\cite{deBoer:2013gz,deBoer:2014fra}.

On the other hand one disadvantage of our approach is that the first term in the on-shell action~\eqref{se3:onshellfinal} depends on the location of the horizon. The existence of a horizon is a non-trivial requirement in higher spin theories since higher spin gauge transformations act non-trivially on the metric~\cite{Gutperle:2011kf,Castro:2011fm,Ammon:2011nk}. Hence the latter can remove or restore the horizon of the higher spin black hole. Nevertheless we may assume that it is always possible to find a gauge where the horizon is manifest and the on-shell action~\eqref{se3:onshellfinal} makes sense. In particular note that the metric obtained from the Chern-Simons connections given in eqs.~\eqref{se3:sources1} and~\eqref{se3:sources2} does not have an event horizon. This will not be a problem for the higher spin black hole of Ammon, Gutperle, Kraus, and Perlmutter~\cite{Gutperle:2011kf,Ammon:2011nk} since the gauge transformation that restores the horizon was found in~\cite{Ammon:2011nk}.


\section{The AGKP black hole} 
\label{se4:thermodynamics}

Let us now consider the thermodynamics of the higher spin black hole of Ammon, Gutperle, Kraus, and Perlmutter~\cite{Gutperle:2011kf,Ammon:2011nk}. The black hole is described in the wormhole gauge by the connections
  \eq{
  a &= \( L_{+1} - \frac{2\pi}{k}\L \,L_{-1} - \frac{\pi}{2k}\W \,W_{-2} \)\, dx^+ \notag \\
   &+ \mu \lb W_{+2} +\frac{4\pi}{k}\W\,L_{-1} - \frac{4\pi}{k}\L\,W_0 + \frac{4\pi^2}{k^2} \L^2 W_{-2}  \rb dx^-,   \label{se4:agkp1} \\
  \bar{a}& = -\( L_{-1} - \frac{2\pi}{k} \bar{\L} \,L_{+1} - \frac{\pi}{2k}\overline{\W} \, W_{+2} \) dx^- \notag \\
  &-\bar{\mu} \lb W_{-2} + \frac{4\pi}{k} \overline{\W}\,L_{+1} -\frac{4\pi}{k} \bar{\L}\,W_0 + \frac{4\pi^2}{k^2}\bar{\L}^2 W_{+2} \rb dx^+.  \label{se4:agkp2}
  }
These solutions are closely related to those given in eqs.~\eqref{se3:sources1} and~\eqref{se3:sources2} except that they do not satisfy all the conditions of the Fefferman-Graham gauge.  Indeed, while the connections above obey eq.~\eqref{se3:fgcsextra}, they only satisfy part of eq.~\eqref{se3:fgcs3} since
  \eq{
  \tr \lb (A-\bar{A})_{\mu}\, W_0 \rb \ne 0. \label{se4:brokenfg}
  }

Eq.~\eqref{se4:brokenfg} implies that we cannot use the boundary term~\eqref{se3:csboundaryfinal} when evaluating the action on shell. Indeed, away from the Fefferman-Graham gauge we expect the appropriate boundary terms accompanying the action of higher spin theories to be highly non-linear in the Chern-Simons fields. One may attempt to put the AGKP solution in the Fefferman-Graham gauge via an $SL(3,R)$ gauge transformation. This is not possible, however, since the resulting solution will not be compatible with the boundary conditions given in eqs.~\eqref{se3:sourcebc1} and~\eqref{se3:sourcebc2}. The moral of the story is that the AGKP black hole obeys more restrictive boundary conditions than those admissible by the Fefferman-Graham gauge (cf.~the discussion above eq.~\eqref{se3:nonfgbc}) and the asymptotic boundary terms found in the previous section are not applicable in this case.

Nevertheless we will assume that the on-shell action of the AGKP black hole is still given by eq.~\eqref{se3:onshellfinal}. We should then note that the solution described by eqs.~\eqref{se4:agkp1} and~\eqref{se4:agkp2} does not have a horizon and may be interpreted as a wormhole connecting two asymptotically-AdS$_3$ solutions~\cite{Gutperle:2011kf,Ammon:2011nk}. Finding the gauge transformation that makes the horizon manifest is a technically challenging task which was accomplished in~\cite{Ammon:2011nk} for a black hole with parameters
  \eq{
  \bar{\L} = \L, \quad \overline{\W} = - \W, \quad \bar{\mu} = - \mu.
  }
The black hole gauge, where the horizon of the higher spin black hole is manifest, is described by connections $\A$ and $\bar{\A}$ related to the wormhole gauge via a gauge transformation
  \eq{
  \A = U^{-1} A U + U^{-1} d U, \\
  \bar{\A} = U \bar{A} U^{-1} + U d U^{-1},
  }
where $A$ and $\bar{A}$ are given in terms of $a$ and $\bar{a}$ by eq.~\eqref{se2:fgcs} and the $SL(3,R)$ element $U$ is given by
  \eq{
  U = e^{F(\xi)(W_{+1} - W_{-1}) + G(\xi) L_0}, \qquad \xi = \log \(\frac{r}{r_\H}\).
  }
The functions $F(\xi)$ and $G(\xi)$ are given in eq.~(D.10) of ref.~\cite{Ammon:2011nk} and $r_\H$ is the location of the horizon. The latter agrees with that of the BTZ black hole~\eqref{se2:horizon} which is the solution the AGKP black hole reduces to in the limit $\W \ra 0$, $\mu \ra 0$. 

In analogy with the BTZ black hole, the temperature and chemical potential of the AGKP black hole may be determined from (a) the identification $t_E \sim t_E + \b$ that guarantees a smooth horizon free of conical singularities in the Euclidean continuation of the metric and higher spin field; or (b) by demanding trivial holonomies for the Chern-Simons connections around the contractible time cycle. The inverse temperature and chemical potential are given by~\cite{Gutperle:2011kf,Ammon:2011nk}
  \eq{
  \b&= \pi\sqrt{\frac{k}{2\pi\L}} \,\frac{2c-3}{c-3}\,\frac{\sqrt{c}}{\sqrt{4c-3}}, \qquad \mu = \frac{3}{4} \sqrt{\frac{k}{2\pi\L}} \,\frac{\sqrt{c}}{2c-3}, \label{se4:chemicalpotentials}
  }
where the constant $c$ is defined via
  \eq{
  \frac{c-1}{c^{3/2}} = \sqrt{\frac{k}{32\pi}\frac{\W^2}{\L^3}}.
  }

Since the Chern-Simons actions $I_{CS}[\A]$ and $I_{CS}[\bar{\A}]$ vanish on shell, the covariant, on-shell action of the AGKP black hole is given by
  \eq{
  I_3^{os}[\A,\bar{\A}] = \frac{k_3}{4\pi} \int_{\H} \tr \( \A \we \bar{\A} \) - 2\dsint \mu \W\, dtd\phi.\label{se4:onshellfinal}
  }
While we did not derive the asymptotic boundary term in eq.~\eqref{se4:onshellfinal} we have conjectured that it takes the same form as in eq.~\eqref{se3:onshellfinal}. We can confirm that this is indeed the case by studying the thermodynamics of the AGKP black hole.
  
The free energy of the AGKP black hole that follows from eq.~\eqref{se4:onshellfinal} is given by
  \eq{
  \F &= - \frac{1}{\b} \frac{k_3}{4\pi} \int_{\H} \tr \( \A \we \bar{\A} \) + 4\pi \mu \W, \\
  \F &= - 4\pi\L\,\frac{ 4 c^3 - 21 c^2 + 24 c - 9}{c\( 3 - 2c \)^2},
  }
where once again we have used $t \ra -it_E$ in the Euclidean continuation of the action. In the canonical ensemble the temperature and chemical potential are held fixed and the free energy is a function of these variables. Indeed, for large temperature/small chemical potential we have
  \eq{
  \F = -\frac{2\pi^2 k}{\b^2} \( 1 + \frac{32 \pi^2}{3} \frac{\mu^2}{\b^2} + \frac{5120 \pi^4}{27} \frac{\mu^4}{\b^4} + \dots \),
  }
in agreement with the free energy of the BTZ black hole~\eqref{se2:freeenergy} in the limit $\mu \ra 0$.
    
  %
  %
  
From the free energy it is easy to compute the entropy,
  \eq{
  S = \b^2 \frac{\p\F}{\p\b}  = 4\pi \sqrt{2\pi k \L} \(1-\frac{3}{2c}\)^{-1} \sqrt{1-\frac{3}{4c}},
  }
which agrees with the result obtained via Wald's entropy formula~\cite{Campoleoni:2012hp}, by integrating the first law of thermodynamics~\cite{Perez:2013xi}, and in the microcanonical ensemble~\cite{Banados:2012ue,deBoer:2013gz} (see~\cite{Ammon:2012wc,Perez:2014pya} for reviews and references on other approaches). Note that it was not necessary to express the free energy, that is the variables $c$ and $\L$, in terms of the temperature and chemical potentials in order to derive the entropy. Indeed, it is sufficient to find their derivatives with respect to $\b$ and $\mu$ via eq.~\eqref{se4:chemicalpotentials}. In particular, it was not necessary to find the charges of the black hole in order to determine its entropy. These charges are non-linear expressions in $\mu$, $\L$, and $\W$~\cite{Compere:2013gja,Compere:2013nba}.

From the free energy we can also determine the charges conjugate to the temperature and chemical potential. The total energy is given by 
  %
  %
  \eq{
  \cE_T = \frac{\p (\b\F)}{\p\b} & = 4\pi \( \L+\mu \W - \frac{32\pi}{3k} \mu^2 \L^2 \),
  }
which agrees with the charge computed in the canonical formalism~\cite{Perez:2012cf} provided that variations of $\mu$ do not contribute to variation of the charge. Not surprisingly the charge conjugate to the chemical potential is proportional to the spin-3 charge of the undeformed theory,
  \eq{
  \J = -\frac{\p \F}{\p\mu} = 8\pi \W.
  }
In terms of these variables the free energy reads,
  \eq{
  \F = \cE - \mu \J - \cS/\b, 
  }  
where $\cE_T = \cE - \mu \J$. This expression is compatible with a partition function of the form,
  \eq{
  Z(\b,\mu) = \tr \lb e^{-\b \(E - \mu J \)} \rb,
  }
which is what we expect to find in the canonical formalism and the covariant approach to the thermodynamics of black holes. 

We have seen that the on-shell action given in eq.~\eqref{se4:onshellfinal} captures the thermodynamics of the AGKP black hole. Thus, when $\overline{\W} = -\W$ and $\bar{\mu} = -\mu$, both the solution considered in the previous section, cf.~eqs.~\eqref{se3:sources1} and~\eqref{se3:sources2}, and the AGKP black hole receive a contribution to the on-shell action of the form
  \eq{
  I^{os}_{b(3)} = - 2 \dsint \mu \W\, dtd\phi. \label{se4:boundaryterm}
  }
When (non-constant) sources for the higher spin charges are turned on we expect to obtain the induced action of $\W$-gravity at the boundary. Computing the induced action of $\W$-gravity falls outside the scope of this paper. Instead we will assume that the on-shell value of the asymptotic boundary term captures some of the properties of this action. 
  
In contrast to the solution considered in the previous section the charges of the AGKP solution do obey the chiral $\W_3$ Ward identities described by eqs.~\eqref{se3:ward1} and~\eqref{se3:ward2}.\footnote{Note that all parameters of the AGKP solution are constant. To see that the charges obey the chiral Ward identities one needs to consider the non-constant solutions given in~\cite{Gutperle:2011kf}.} The AGKP solution achieves this by obeying more restrictive boundary conditions than those admissible by the Fefferman-Graham gauge. In particular the spin-3 field satisfies, cf.~eq.~\eqref{se3:nonfgbc},
  \eq{
  \psi_{\mu r r} = \frac{1}{r^2}\, \vp_{\mu r r} + \textrm{subleading terms}, \label{se4:vector}
  }
where $\vp_{\mu r r}$ depends only on the boundary coordinates.

In the previous section we argued that the charges in the boundary term~\eqref{se4:boundaryterm} do not obey the chiral Ward identities due to general covariance of the induced action of $\W$-gravity. This suggests that the boundary conditions obeyed by the AGKP black hole lead to a different theory of induced gravity at the boundary. From the arguments presented in the previous section one may think that this action is not covariant. This cannot be the case, however, since the higher spin theory is itself covariant. Indeed, in contrast with eq.~\eqref{se3:imtired}, variation of the action is now given by
  \eq{
  \d I_3[A,\bar{A}] = \dsint  \Big\{ (\W + \pi_1)\, \d \mu + (\overline{\W} + \pi_2 ) \, \d \bar{\mu} + \pi_3\,\d \vp_{-rr} + \pi_4 \d \vp_{+rr} \Big \} d^2x, \label{se4:newvariation}
  }
where the functions $\pi_{i = 1\dots 4}$ are expected to be non-linear functions of the charges and sources. Eq.~\eqref{se4:newvariation} reflects the fact that additional boundary conditions are fixed for the AGKP black hole. 


\section{Conclusions}
\label{se:conclusions}

In this paper we have proposed a set of boundary terms for higher spin theories in AdS$_3$ that lead to a finite action with a well-defined variational principle compatible with Dirichlet boundary conditions for the metric and higher spin fields. We expect the boundary terms to be uniquely determined by general covariance of the second order formulation of the theory which suggests that they are highly non-linear expressions of the Chern-Simons fields. Nevertheless we argued that in the Fefferman-Graham gauge the boundary terms become quadratic expressions of the latter. We showed this explicitly for the $SL(2,R)$ theory and conjectured it to hold for higher spin theories as well.

We showed that in Schwarzschild coordinates the covariant on-shell action receives a contribution from a generalization of the Gibbons-Hawking term at the horizon. This term arises in the derivation of the Chern-Simons formulation from the first order formulation of the theory. Keeping this boundary term in the action was equivalent to working in the first order formulation of the higher spin theory but the Chern-Simons language proved to be more convenient. Although the existence of a horizon is not a gauge-invariant concept, it was comforting to find that, when the horizon is manifest, its contribution to the on-shell action led to the correct thermodynamics of the AGKP black hole. 

Our results suggest that the AGKP boundary conditions do not lead to the standard theory of induced $\W$-gravity at the boundary. Note, however, that the AGKP boundary conditions are not related to the standard boundary conditions (plus sources) of~\cite{Campoleoni:2010zq} via a gauge transformation. Hence it is not surprising that theories obeying different boundary conditions in the bulk lead to different induced theories at the boundary.

While we did not discuss black holes in the diagonal embedding~\cite{Castro:2011fm} it should be straightforward to generalize our results to that setting. In particular let us point out that, although the solutions presented in~\cite{Castro:2011fm} are not in the Fefferman-Graham gauge, their chemical potential and temperature depend only on the lower spin charge and spin-2 charge, respectively. This implies that the entropy only receives contributions from the generalized Gibbons-Hawking term at the horizon, as in the BTZ case.

Let us conclude by suggesting future directions for the present work. First of all it would be interesting to use our approach to study the thermodynamics of the solutions presented in eqs.~\eqref{se3:sources1} and~\eqref{se3:sources2}. Although the holonomy conditions that yield a smooth solution in Euclidean signature are more complicated than those of the AGKP black hole, the main obstacle is finding a gauge where the event horizon is manifest. It would also be interesting to extend our boundary terms to the metric formulation of the $SL(3,R)$ higher spin theory studied in~\cite{Fujisawa:2012dk,Campoleoni:2012hp} or to derive these by demanding a well-defined variational principle of the second order action. 

Finally it would be interesting to relax some of the conditions we have imposed on the action. For example refs.~\cite{Arcioni:2002vv,Apolo:2015fja} considered boundary terms that lead to a well-defined variational principle (in the second order formulation of three-dimensional gravity) only on shell. In~\cite{Apolo:2015fja} it was shown that, when accompanied by free boundary conditions~\cite{Compere:2008us,Compere:2013bya,Troessaert:2013fma,Avery:2013dja,Apolo:2014tua}, the WZW model describing the asymptotic dynamics of three-dimensional gravity~\cite{Coussaert:1995zp} is supplemented by the Virasoro constraints of string theory. We expect that a similar approach in higher spin theories leads to the action of so-called $\W$-strings at the boundary which result from gauging the higher spin sources~\cite{Hull:1993kf}. A set of boundary conditions relevant in this context were recently proposed in~\cite{Poojary:2014ifa}.


\section*{Acknowledgments}

It is a pleasure to thank Bo Sundborg and Nico Wintergerst for helpful discussions, and Bo Sundborg for comments on the manuscript. I am also grateful to the anonymous referee for their valuable comments.

\appendix

\section{Conventions}
\label{ap:conventions}

In the main text we have used the following representation of $SL(2,R)$
  \def\matrixsize{14pt}
  \eq{
  L_{-1} = \( \begin{array}{*2{C{\matrixsize}}}
			   0 & 1 \\
			   0 & 0
   	 		   \end{array} \), 
\qquad   L_{0} = \frac{1}{2}\( \begin{array}{*2{C{\matrixsize}}}
			   1 & 0 \\
			   0 & -1
   	 		   \end{array} \),
\qquad   L_{+1} = \( \begin{array}{*2{C{\matrixsize}}}
			   0 & 0 \\
			   -1 & 0
   	 		   \end{array} \).
}
These generators satisfy the $SL(2,R)$ algebra
  \eq{
  [L_n, L_m] = (n-m) L_{n+m}.
  }

On the other hand, the representations of $SL(3,R)$ are characterized by how $SL(2,R)$ is embedded in $SL(3,R)$. We work in the principal embedding where a convenient representation is given by
  \eq{
    &L_{-1} = \( \begin{array}{*3{C{\matrixsize}}}
			   0 & -2 & 0 \\
			   0 & 0 & -2 \\
			   0 & 0 & 0
   	 		   \end{array} \), 
&\quad L_{0} = &\( \begin{array}{*3{C{\matrixsize}}}
			   1 & 0 & 0 \\
			   0 & 0 & 0 \\
			   0 & 0 & -1
   	 		   \end{array} \),
 &  L_{+1} =& \( \begin{array}{*3{C{\matrixsize}}}
			   0 & 0 & 0 \\
			   1 & 0 & 0 \\
			   0 & 1 & 0
   	 		   \end{array} \), \notag \\
 &W_{-2} = \( \begin{array}{*3{C{\matrixsize}}}
			   0 & 0 & 8 \\
			   0 & 0 & 0 \\
			   0 & 0 & 0
   	 		   \end{array} \), 
 &\quad W_{-1} =  & \( \begin{array}{*3{C{\matrixsize}}}
			   0 & -2 & 0 \\
			   0 & 0 & 2 \\
			   0 & 0 & 0
   	 		   \end{array} \),
 & W_{0} = \frac{2}{3} &\( \begin{array}{*3{C{\matrixsize}}}
			   1 & 0 & 0 \\
			   0 & -2 & 0 \\
			   0 & 0 & 1
   	 		   \end{array} \), \\
  &W_{+1} = \( \begin{array}{*3{C{\matrixsize}}}
			   0 & 0 & 0 \\
			   1 & 0 & 0 \\
			   0 & -1 & 0
   	 		   \end{array} \), 
 &\quad W_{+2} = &\( \begin{array}{*3{C{\matrixsize}}}
			   0 & 0 & 0 \\
			   0 & 0 & 0 \\
			   2 & 0 & 0
   	 		   \end{array} \).
			\quad   && \notag
  }
These generators satisfy the $SL(3,R)$ algebra
  \eq{
  [L_n, L_m] &= (n-m) L_{n+m}, \qquad [L_n, W_m] = (2n - m) W_{n+m},\\
  [W_n,W_m] & = -\tfrac{1}{3} (n-m)(2n^2 + 2m^2 - nm - 8) L_{n+m}.
  }
  %


\section{Generalization to $SL(N,R)$ higher spin theories} 
\label{ap:sln}

Let us briefly comment on the generalization of the boundary terms to $SL(N,R)$ higher spin theories in the principal embedding. At the asymptotic boundary we have the generalization of the Gibbons-Hawking term that guarantees a well-defined variational principle for the metric and higher spin fields in the first order formulation of the theory. This term introduces divergences to the otherwise finite Chern-Simons action. In the Fefferman-Graham gauge the gauge fields associated with the $SL(2,R)$ part of the algebra, i.e.~$L_{\pm}$, $L_0$, are responsible for divergences of $O(r^2)$. Since these fields are present in all $SL(N,R)$ higher spin theories theories the boundary term is given by
  \eq{
   I_{b(n)} = & \frac{k_n}{4\pi} \dsint \tr \( A\we \bar{A} \) - \frac{k_n}{4\pi}\dsint \tr \lb (A-\bar{A}) \we (A-\bar{A}) L_0 \rb + B_{n}, \label{ap:boundaryterm}
  }
where the level of the $SL(N,R)$ Chern-Simons action is given by~\cite{Campoleoni:2010zq}
  \eq{
  k_n = \frac{k}{2 \,\tr(L_0 L_0)} = \frac{1}{8G_N \tr(L_0\, L_0)},
  }
and $B_n$ denotes additional boundary terms that cancel divergences of $O(r^{2(n-1)})$ associated with the spin-$n$ part of the algebra. In particular, once $\dsint \tr \( A\we \bar{A} \)$ is introduced at the asymptotic boundary we expect the additional boundary terms to be fixed uniquely by requiring cancellation of divergences. Also note that eq.~\eqref{ap:boundaryterm} reduces to the corresponding expression in the $SL(2,R)$ theory, eq.~\eqref{se2:csboundaryfinal}, when the higher spin fields are turned off.

It is clear that, in analogy to the $SL(2,R)$ and $SL(3,R)$ cases, the covariant on-shell action receives contributions from the generalized Gibbons-Hawking term at the horizon. The origin of this term is not mysterious: it arises when deriving the Chern-Simons formulation of higher spin theories from their first order formulation. In particular, since the Chern-Simons action vanishes on shell in the Fefferman-Graham gauge, the covariant action of $SL(N,R)$ black holes is given entirely by the boundary terms
  \eqsp{
  I^{os}_n[A,\bar{A}] = \frac{k_n}{4\pi} \int_{\H} \tr \( A \we \bar{A} \) + I^{os}_{b(n)}[A,\bar{A}].
  }
We expect the on-shell asymptotic boundary term to be given by couplings between the higher spin sources and the corresponding higher spin charges of the undeformed theory. The relationship between these quantities and the Chern-Simons variables is determined from the variation of the action. 


\ifprstyle
	\bibliographystyle{apsrev4-1}
\else
	\bibliographystyle{utphys}
\fi

\bibliography{hsblackholes}

\end{document}